\documentclass[10pt]{article}
\usepackage[final]{graphics}
\usepackage{amsfonts,amsbsy}

\def\empile#1\over#2{\mathrel{\mathop{\kern 0pt#1}\limits_{#2}}}

\newcommand{\sll}{\raise.15ex\hbox{$/$}\kern-.43em\hbox{$l$}}
\newcommand{\slepsilon}{\raise.15ex\hbox{$/$}\kern-.53em\hbox{$\epsilon$}}
\newcommand{\slvarepsilon}{\raise.15ex\hbox{$/$}\kern-.53em\hbox{$\varepsilon$}}
\newcommand{\slL}{\raise.15ex\hbox{$/$}\kern-.53em\hbox{$L$}}
\newcommand{\slP}{\raise.15ex\hbox{$/$}\kern-.53em\hbox{$P$}}
\newcommand{\slp}{\raise.1ex\hbox{$/$}\kern-.63em\hbox{$p$}}
\newcommand{\slq}{\raise.1ex\hbox{$/$}\kern-.63em\hbox{$q$}}
\newcommand{\slv}{\raise.1ex\hbox{$/$}\kern-.63em\hbox{$v$}}
\newcommand{\slR}{\raise.15ex\hbox{$/$}\kern-.53em\hbox{$R$}}
\newcommand{\slQ}{\raise.15ex\hbox{$/$}\kern-.53em\hbox{$Q$}}
\newcommand{\slK}{\raise.15ex\hbox{$/$}\kern-.53em\hbox{$K$}}
\newcommand{\slk}{\raise.15ex\hbox{$/$}\kern-.53em\hbox{$k$}}
\newcommand{\slSigma}{\raise.15ex\hbox{$/$}\kern-.53em\hbox{$\Sigma$}}
\newcommand{\slcalP}{\raise.15ex\hbox{$/$}\kern-.63em\hbox{$\cal P$}}
\newcommand{\slA}{\raise.15ex\hbox{$/$}\kern-.73em\hbox{$A$}}
\newcommand{\slbfA}{\raise.15ex\hbox{$/$}\kern-.73em\hbox{${\imb A}$}}
\newcommand{\slpartial}{\raise.15ex\hbox{$/$}\kern-.53em\hbox{$\partial$}}

\font\tenimbf=cmmib10 at 10pt
\font\sevenimbf=cmmib10 at 7pt
\font\fiveimbf=cmmib10 at 5pt
\newfam\imbf
\textfont\imbf=\tenimbf
\scriptfont\imbf=\sevenimbf
\scriptscriptfont\imbf=\fiveimbf
\def\imb{\fam\imbf\tenimbf}

\def\p{{\boldsymbol p}}
\def\q{{\boldsymbol q}}
\def\l{{\boldsymbol l}}
\def\k{{\boldsymbol k}}
\def\m{{\boldsymbol m}}
\def\x{{\boldsymbol x}}
\def\X{{\boldsymbol X}}
\def\r{{\boldsymbol r}}
\def\z{{\boldsymbol z}}

\def\be{\begin{eqnarray}}
\def\ee{\end{eqnarray}}

\begin{document}

\thispagestyle{empty}
\title {\bf From DIS to proton-nucleus collisions\\ in the Color Glass Condensate model}

\author{Fran\c cois Gelis$^{(1)}$ and Jamal Jalilian-Marian$^{(2)}$}
\maketitle
\begin{center}
\begin{enumerate}
\item Service de Physique Th\'eorique\\
B\^at. 774, CEA/DSM/Saclay\\
91191, Gif-sur-Yvette Cedex, France
\item Physics Department\\
         Brookhaven National Laboratory\\
         Upton, NY 11973, USA
\end{enumerate}
\end{center}

\begin{abstract}

\noindent We show that particle production in proton-nucleus (pA)
collisions in the Color Glass Condensate model can be related to Deep
Inelastic Scattering of leptons on protons/nuclei (DIS).  The common
building block is the quark antiquark (or gluon-gluon) dipole cross
section which is present in both DIS and pA processes.  This
correspondence in a sense generalizes the standard leading twist
approach to pA collisions based on collinear factorization and
perturbative QCD, and allows one to express the pA cross sections in
terms of a universal quantity (dipole cross section) which, in
principle, can be measured in DIS or other processes.  Therefore,
using the parameterization of dipole cross section at HERA, one can
calculate particle production cross sections in proton-nucleus
collisions at high energies. Alternatively, one could use
proton-nucleus experiments to further constrain models of the dipole
cross-section.  We show that the McLerran-Venugopalan model predicts
enhancement of cross sections at large $p_\perp$ (Cronin effect) and
suppression of cross sections at low $p_\perp$. The cross over depends
on rapidity and moves to higher $p_\perp$ as one goes to more forward
rapidities.

\end{abstract}

\vskip 4mm
\centerline{\hfill SPhT-T02/167}

\section{Introduction}

High gluon density effects at high energy (small $x$) have been the
subject of intense theoretical and experimental investigation
\cite{GriboLR1,MuellQ1,McLerV1,McLerV2,McLerV3,Kovch1,Kovch2,JalilKMW1,JalilKLW1,JalilKLW2,JalilKLW3,JalilKW1,AyalaJMV1,AyalaJMV2,KovneM1,KovneMW3,IancuLM1,IancuLM2,IancuM1,FerreILM1}.
At small $x$, the number of gluons per unit area and rapidity in the
wave function of a high energy hadron or nucleus becomes large and the
high energy hadron or nucleus can be described by a classical color
field. Since most of the gluons in the wave function of a high energy
hadron or nucleus are in a coherent state characterized by their
typical momentum $Q_s$, a high energy hadron or nucleus is dubbed a
Color Glass Condensate.  An effective action and renormalization group
approach has been developed which enables one to calculate cross
sections in a high gluon density environment.

In \cite{DumitJ1,DumitJ2,GelisJ1,GelisJ2}, for proton-nucleus
collisions at high energies and in the forward rapidity
region\footnote{By forward rapidity region we mean, roughly, any
  rapidity between proton rapidity and mid-rapidity.}  we proposed
describing the proton by the QCD parton model and the nucleus by the
Color Glass Condensate model. One can then relate particle production
in proton-nucleus collisions to the scattering of a quark or gluon
from the Color Glass Condensate
\cite{DumitJ1,DumitJ2,GelisJ1,GelisJ2}. Here we show that one can
relate particle production in proton-nucleus collisions to Deep
Inelastic Scattering of electrons (or more precisely, virtual photons)
on protons and nuclei. This is due to the fact that the description of
the target as a Color Glass Condensate is universal and independent of
the process considered.

While the relation between particle production in proton-nucleus
collisions and DIS structure functions has been known for a while in
the case of dilepton\footnote{We would like to thank Y. Kovchegov for
pointing this out to us.} (virtual photon) production
\cite{KopelRT1,KopelRTJ1} (and references therein), here we show that
this relation is more general in our framework and holds for
production of any particle (which has a known fragmentation function)
in proton-nucleus collisions at high energy. This is crucial in view of
the upcoming proton (deuteron)-nucleus experiments at RHIC. One would
like to have predictions from QCD with as few model dependent assumptions 
about nuclear effects as possible.

In proton-proton collisions, standard calculations are based on
collinear factorization theorems, proven at the leading twist level.
The predictability of the theory is due to the fact that the building
blocks (distribution or fragmentation functions) used in the cross 
section are universal (i.e. process independent) objects which can be 
measured in a given process and used to predict the outcome of other 
processes. It is well known that higher twist effects break collinear 
factorization and one has to resort to other methods. In proton-nucleus
collisions, in order to take nuclear modifications such as shadowing and 
Cronin effect into account, one commonly modifies the parton distributions to 
include nuclear shadowing and introduces some model for Cronin effect.
The parameters of the model are then constrained by fits to the existing data
and used to make predictions for other processes or higher energies.

However, it is clear that this approach cannot be true in general
since collinear factorization theorems break down due to higher twist
(multiple scatterings) effects which become significant in scattering
off nuclei.  Therefore there is no guarantee that the parameters of
such models are universal and process independent and these models
have little predictive power. Here we show that one can generalize the
standard leading twist expressions in such a way that all higher twist
effects are included and the universality of the building blocks
(dipole cross sections) of the theory is preserved and the predictive
power of the theory is restored.

\section{Reminder: DIS and the dipole cross-section}
\subsection{DIS amplitude}
\begin{figure}[htbp]
\begin{center}
\resizebox*{!}{2cm}{\includegraphics{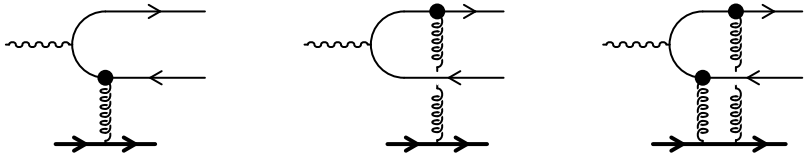}}
\end{center}
\caption{\label{fig:DIS} The relevant diagrams 
  for $\gamma^*A\to q\bar{q}X$ in the Color Glass Condensate model.
  The black dots denote the eikonal interaction between the quark or
  antiquark with the classical color field.}
\end{figure}

We first start by a brief reminder of how the dipole cross-section
appears in Deep Inelastic Scattering. This calculation will in fact be
useful in order to compare with the very similar problem of dilepton
(i.e. virtual photon) production in pA collisions. The nucleus is
treated using the Color Glass Condensate model, i.e. as a set of
classical and randomly distributed color sources that generate a
classical field.  There are three relevant diagrams for the DIS
amplitude in this model, which are represented in figure
\ref{fig:DIS}.
In this figure, the black dot denotes the eikonal all-orders
interaction between a quark or antiquark with the classical field of
the nucleus. More precisely, for a quark or antiquark line in which
the incoming momentum is $P$ and the outgoing momentum\footnote{We use
the following notations throughout this paper: $P$ denotes a
4-momentum, ${\p}$ denotes its spatial part, ${\p_\perp}$ its
transverse components, and $p^\pm$ its longitudinal components in
light-cone variables.} is $Q$, the (time-ordered) eikonal scattering
amplitude reads
\begin{equation}
T_{\rm eik}(Q,P)=2\pi\delta(q^--p^-)\gamma^- {\rm sign}(p^-)
\int d\x_\perp e^{i(\q_\perp-\p_\perp)\cdot \x_\perp}
\left(U^{{\rm sign}(p^-)}(\x_\perp)\!-1\!\right)\; ,
\label{eq:Teik}
\end{equation}
where we assume that the nucleus is moving close to the speed of light
in the $+z$ direction. $U(\x_\perp)$ is a matrix living in the
fundamental representation of $SU(N_c)$, given by
\begin{equation}
U({\imb x}_\perp) \equiv {\rm T} \exp \bigg \{-ig^2 \int^{+\infty}_{-\infty}
d z^- {1 \over {\nabla^2_\perp}} \rho_a (z^-,{\imb z}_\perp) t^a \bigg\}\; ,
\label{eq:Udef}
\end{equation}
with $t^a$ in the fundamental representation, and where
$\rho_a(z^-,{\imb z}_\perp)$ is the density of color sources in the
nucleus. How to average over these sources is explained in
\cite{GelisP1,GelisJ1}.

We first compute the $\gamma^*A\to q\bar{q}X$ amplitude, and bring it
to a form that will be easy to compare with the photon production
amplitude. A straightforward use of the CGC rules (see
\cite{GelisP1,GelisP2,GelisJ1,GelisJ2} for details and examples) gives
for this amplitude the following sum of three terms\footnote{The
  factor $2\pi\delta(k^--p^--q^-)$, which expresses the fact that the
  problem is invariant under translations in $x^+$ has been excluded
  from the definition of the amplitude ${\cal M}^\mu$.}, each
corresponding to a diagram in figure \ref{fig:DIS}:
\begin{eqnarray}
&&\!\!\!\!{\cal M}^\mu_{_{DIS}}(\k|\q,\p)= 
-i\int d^2\x_{2\perp} e^{i(\q_\perp+\p_\perp-\k_\perp)\cdot\x_{2\perp}}
\left(U^\dagger(\x_{2\perp})-1\right)\nonumber\\
&&\qquad\qquad\times\overline{u}(\q)\left[
\frac{\gamma^\mu(\slQ-\slK+m)\gamma^-}{(Q-K)^2-m^2+i\epsilon}
\right]v(\p)\nonumber\\
&&+i\int d^2\x_{1\perp} e^{i(\q_\perp+\p_\perp-\k_\perp)\cdot\x_{1\perp}}
\left(U(\x_{1\perp})-1\right)\nonumber\\
&&\qquad\qquad\times\overline{u}(\q)\left[
\frac{\gamma^-(\slK-\slP+m)\gamma^\mu}{(P-K)^2-m^2+i\epsilon}
\right]v(\p)\nonumber\\
&&+i\int\frac{d^2\l_\perp}{(2\pi)^2}
\int d^2\x_{1\perp}d^2\x_{2\perp}
\nonumber\\
&&\,\;
\times e^{i\l_\perp\cdot\x_{1\perp}}
e^{i(\p_\perp+\q_\perp-\k_\perp-\l_\perp)\cdot\x_{2\perp}}
\left(U(\x_{1\perp})-1\right)\left(U^\dagger(\x_{2\perp})-1\right)
\nonumber\\
&&\;\,\times
\overline{u}(\q)\!\left[
\frac{\gamma^-(\slQ-\slL+m)\gamma^\mu(\slQ-\slK-\slL+m)\gamma^-}%
{2p^-[(\q_\perp\!-\!\l_\perp)^2+m^2\!-\!2q^-k^+]
+2q^-[(\q_\perp\!-\!\k_\perp\!-\!\l_\perp)^2+m^2]}
\right]\!v(\p)\; .\nonumber\\
&&
\label{eq:DIS1}
\end{eqnarray}
In this equation $Q, P$ and $K$ are the momenta of the outgoing quark,
outgoing antiquark and incoming photon respectively, $\x_{1\perp}$ and
$\x_{2\perp}$ are the coordinates in impact parameter space of the
quark and antiquark. In the third diagram, the momentum $L$ is the
momentum transferred between the nucleus and the quark line. Its $-$
component is zero because the nucleus is moving at the speed of light
in the $+z$ direction, and its $+$ component has already been
integrated out by using the theorem of residues (we pick the pole in
the upper half-plane, at
$l^+=q^+-[(\q_\perp-\l_\perp)^2+m^2-i\epsilon]/2q^-$).

At this point, it is useful to note that since $P$ and $Q$ are
on-shell, we have the following two identities:
\begin{eqnarray}
&&\overline{u}(\q)=\frac{1}{2q^-}\overline{u}(\q)\gamma^-(m+\slQ)\; ,
\nonumber\\
&&v(\p)=-\frac{1}{2p^-}(m-\slP)\gamma^-v(\p)\; .
\label{eq:dirac-alg1}
\end{eqnarray}
Inserting them respectively in the first and second term of
Eq.~(\ref{eq:DIS1}), and introducing also a dummy variable $\l_\perp$
via $\int d^2\x_{1\perp} d^2{\l_\perp}/(2\pi)^2 \exp(i \l_\perp\cdot
\x_{1\perp})$ or $\int d^2\x_{2\perp} d^2{\l_\perp}/(2\pi)^2 \exp(i
\l_\perp\cdot \x_{2\perp})$, we can combine three terms into one:
\begin{eqnarray}
&&{\cal M}^\mu_{_{DIS}}(\k|\q,\p)
=\frac{i}{2}\int\frac{d^2\l_\perp}{(2\pi)^2}
\int d^2\x_{1\perp}d^2\x_{2\perp}
\nonumber\\
&&\qquad\;
\times e^{i\l_\perp\cdot\x_{1\perp}}
e^{i(\p_\perp+\q_\perp-\k_\perp-\l_\perp)\cdot\x_{2\perp}}
\left(U(\x_{1\perp})U^\dagger(\x_{2\perp})-1\right)\;
\nonumber\\
&&\qquad\;\times\overline{u}(\q)\,
\Gamma^\mu(k^\pm,\k_\perp|q^-,p^-,\q_\perp-\l_\perp)\,v(\p)
\; ,
\label{eq:DIS2}
\end{eqnarray}
where we denote
\begin{eqnarray}
&&
\Gamma^\mu(k^\pm,\k_\perp|q^-,p^-,\q_\perp-\l_\perp)\equiv\nonumber\\
&&\qquad\equiv
\frac{\gamma^-(\slQ-\slL+m)\gamma^\mu(\slQ-\slK-\slL+m)\gamma^-}%
{p^-[(\q_\perp\!-\!\l_\perp)^2+m^2\!-\!2q^-k^+]
+q^-[(\q_\perp\!-\!\k_\perp\!-\!\l_\perp)^2+m^2]}
\; .
\label{eq:commonspinors}
\end{eqnarray}
This is our final expression for the DIS amplitude off the target
color field. Note that $\Gamma^\mu$ a priori depends on $\p_\perp$ and
$\q_\perp$. However, in its list of arguments, we have anticipated the
fact that the $\p_\perp$ and $\q_\perp$ that comes from the spinors
drops out in physical quantities, because in the squared amplitude
they always appear in combinations such as $\gamma^-
u(\q)\overline{u}(\q)\gamma^-=2q^-\gamma^-$. In Eq.~(\ref{eq:DIS2}),
$\Gamma^\mu$ is a quantity which describes how the virtual photon
splits into a $q\bar{q}$ pair, which longitudinal momenta $q^-,p^-$
and transverse momentum of the quark $\q_\perp-\l_\perp$, while the
factor $U(\x_{1\perp})U^\dagger(\x_{2\perp})-1$ can be seen as the
scattering amplitude of the dipole on color field of the nucleus.

\subsection{DIS Cross-section}
{}From Eq.~(\ref{eq:DIS2}), one can obtain the $\gamma^*A$ cross-section
as
\begin{eqnarray}
&&
d\sigma_{_{DIS}}=\frac{d^3\q}{(2\pi)^22q_0}\frac{d^3\p}{(2\pi)^32p_0}
\frac{1}{2k^-} 
2\pi\delta(k^--p^--q^-)\nonumber\\
&&\qquad\qquad\times
\left<{\cal M}^\mu_{_{DIS}}(\k|\q,\p){\cal M}^{\nu *}_{_{DIS}}(\k|\q,\p)
\right>_\rho
\epsilon_\mu(K)\epsilon_\nu^*(K)\; ,
\end{eqnarray}
where $\left<\cdots\right>_\rho$ denotes the average over the color
sources and where $\epsilon^\mu(K)$ is the polarization vector of the
photon.  At this point, in order to exploit the fact that $\Gamma^\mu$
depends only on $\m_\perp\equiv\q_\perp-\l_\perp$, we can use this
quantity as the integration variable in ${\cal M}^\mu_{_{DIS}}$, and
we can integrate out trivially the transverse momenta $\q_\perp$ and
$\p_\perp$ of the quark and antiquark in the final state, in order to
obtain\footnote{We have made explicit the fact that it is the real
part of the correlator $\left<\cdots\right>_\rho$ which appears in the
total cross-section, in order to emphasize the connection with the
optical theorem. However, the real part has no effect since this
correlator is purely real in the Color Glass Condensate
model. Therefore, we drop it in the following formulas.}:
\begin{eqnarray}
&&\sigma_{_{DIS}}=\frac{1}{32\pi k_-^2}\int_0^1 \frac{dz}{z(1-z)}
\int d^2\r_\perp 
\int \frac{d^2\m_\perp}{(2\pi)^2}\frac{d^2\m^\prime_\perp}{(2\pi)^2}
\,\epsilon_\mu(k)\epsilon_\nu^*(k)
\nonumber\\
&&\!\!\!\!\!\!\!\!\times
e^{i(\m_\perp-\m^\prime_\perp)\cdot\r_\perp}
\int d^2\X_\perp\;{\rm Tr}_c
\left(1-{\rm Re}\left<U(\X_\perp+\frac{\r_\perp}{2})
U^\dagger(\X_\perp-\frac{\r_\perp}{2})\right>_\rho\right)
\nonumber\\
&&\!\!\!\!\!\!\!\!\times
{\rm Tr}_d\left((\slq+m)
\Gamma^\mu(k^\pm,\k_\perp|q^-,p^-,\m_\perp)(\slp-m)
\Gamma^{\nu\dagger}(k^\pm,\k_\perp|q^-,p^-,\m^\prime_\perp)
\right)\; ,
\end{eqnarray}
with $z$ the momentum fraction $z\equiv q^-/k^-$ ($p^-=(1-z)k^-$),
where ${\rm Tr}_c$ denotes the color trace and ${\rm Tr}_d$ the Dirac
trace, and where we have introduced the dipole size
$\r_\perp\equiv\x_{1\perp}-\x_{2\perp}$ and barycenter $\X_\perp\equiv
(\x_{1\perp}+\x_{2\perp})/2$. If we introduce the dipole
cross-section\footnote{Strictly speaking, the dipole cross-section
should depend on the longitudinal momentum fractions $z$ and $1-z$ of
the quark and antiquark, because these parameters control the rapidity
interval between the quark or antiquark and the target.  However, the
small $x$ evolution that we will consider later aims at resumming
leading powers of $\ln(1/x)$, which means that we neglect $\ln(1/z)$
and $\ln(1/(1-z))$ in front of $\ln(1/x)$. This is a reasonable
approximation at high energy (small $x$) and if the photon wave
function does not overweight values of $z$ close to $0$ or $1$.}
\begin{equation}
\sigma_{\rm dipole}(\r_\perp)\equiv \frac{2}{N_c}
\int d^2\X_\perp\;{\rm Tr}_c
\left<1-U(\X_\perp+\frac{\r_\perp}{2})
U^\dagger(\X_\perp-\frac{\r_\perp}{2})\right>_\rho
\label{eq:sigma-dipole}
\end{equation}
and the square of the photon wave function:
\begin{eqnarray}
&&\left|\Psi(k^\pm,\k_\perp|z,\r_\perp)\right|^2\equiv
\frac{N_c\,\epsilon_\mu(K)\epsilon_\nu^*(K)}{64\pi k_-^2 z(1-z)}
\int \frac{d^2\m_\perp}{(2\pi)^2}\frac{d^2\m^\prime_\perp}{(2\pi)^2}
e^{i(\m_\perp-\m^\prime_\perp)\cdot\r_\perp}
\nonumber\\
&&\quad\times
{\rm Tr}_d\left((\slq+m)
\Gamma^\mu(k^\pm,\k_\perp|q-,p^-,\m_\perp)(\slp-m)
\Gamma^{\nu\dagger}(k^\pm,\k_\perp|q^-,p^-,\m^\prime_\perp)
\right)\, ,\nonumber\\
&&
\end{eqnarray}
we can write the following well known formula for the $\gamma^*A$
cross-section:
\begin{equation}
\sigma_{_{DIS}}=\int_0^1 dz\int d^2\r_\perp
\left|\Psi(k^\pm,\k_\perp|z,\r_\perp)\right|^2
\sigma_{\rm dipole}(\r_\perp)\; .
\label{eq:DIScross-section}
\end{equation}
So far, we have treated the scattering of the quark and antiquark off
the nucleus at a purely classical level, and disregarded its energy
dependence. This energy dependence is usually encoded in the
dependence upon the variable $X\equiv - K^2/2K\cdot P_n$ where $P_n$ is
the 4-momentum of a nucleon inside the nucleus\footnote{At leading
  twist, and in the frame where the DIS can be seen as a quark of the
  nucleus being struck by the virtual photon, $x$ has the
  interpretation of the longitudinal momentum fraction of the struck
  quark.}.  In the CGC model, this energy dependence is expected to
arise in the functional which is used in order to perform the average
over the sources, when one solves the renormalization group equation
that controls its $x$-dependence. 

\subsection{Models for the dipole cross-section}
\subsubsection{McLerran-Venugopalan model}
A first possibility to estimate the dipole cross-section is to use the
McLerran-Venugopalan model, i.e. to use a Gaussian distribution of the
sources $\rho_a(z^-,{\z_\perp})$ that contribute to the matrix
$U(\x_\perp)$ defined in Eq.~(\ref{eq:Udef}). Assuming a distribution
of the form
\begin{equation}
W[\rho]\equiv\exp\left\{
-\int dx^- d^2\x_\perp \frac{\rho_a(x^-,\x_\perp)
\rho^a(x^-,\x_\perp)}{2\mu^2(x^-,\x_\perp)}
\right\}\; ,
\end{equation}
with $\mu^2(x^-,\x_\perp)$ a density of color sources per unit volume,
one obtains (see the appendix A of \cite{GelisP1}):
\begin{eqnarray}
\sigma_{\rm dipole}(\r_\perp)=\pi R^2\left[1-\exp\left(-Q_s^2\int d^2\z_\perp
[G_0(z_\perp)-G_0(\z_\perp-\r_\perp)]^2
\right)\right]\; ,
\label{eq:dipMV}
\end{eqnarray}
where $Q_s^2\sim \alpha_s^2\int dz^- \mu^2(z^-)$ (in order to arrive
at this formula for the dipole cross-section, one has to assume that
the target is homogeneous in the transverse plane) and where $G_0$ is the free propagator in two dimensions:
\begin{equation}
G_0(\z_\perp)=\int \frac{d^2\k_\perp}{(2\pi)^2}
\frac{e^{i \k_\perp\cdot \z_\perp}}{\k_\perp^2}
=\frac{1}{4\pi}\ln\left(\frac{1}{\z_\perp^2 \Lambda^2}\right)\; ,
\end{equation}
where $\Lambda$ is some infrared cutoff related to the scale at which
color neutrality occurs. Therefore, $\Lambda$ is at least as large as
the inverse hadron size, i.e. $\Lambda_{_{QCD}}$. In fact, it can be
proven that in a saturated target, color neutrality occurs over
transverse spatial scales as small as $Q_s^{-1}$
\cite{IancuM1,FerreIIM1}.  At small dipole sizes, this integral
behaves like $r_\perp^2 \ln(1/r_\perp\Lambda)$.  Quantum corrections
to the McLerran-Venugopalan model give an $x$ dependence to the
saturation scale $Q_s$ and hence to the dipole cross-section
\cite{JalilKLW1,JalilKLW2,IancuLM1,IancuLM2,IancuM1,FerreILM1,Muell3,Kovch5}.

\subsubsection{Golec-Biernat-W\"usthoff model - without evolution}
In \cite{GolecW1,GolecW2,GolecW3}, Golec-Biernat and W\"usthoff
developed a model of dipole cross-section that is loosely inspired
by the previous form of the dipole cross-section:
\begin{equation}
\sigma_{\rm dipole}(x,\r_\perp)=\sigma_0(1-\exp(-r_\perp^2/4R_0^2(x)))\; ,
\label{eq:dipGBW}
\end{equation}
with $R_0^2(x)[{\rm GeV}^{-2}]=(x/x_0)^\lambda$.  The main
simplification compared to Eq.~(\ref{eq:dipMV}) is that the $\r_\perp$
dependence is taken to be strictly Gaussian, i.e. one neglects the
slowly varying logarithm in the exponential\footnote{When we connect
particle production in $pA$ collisions to the dipole cross-section in
section \ref{sec:pA}, we will show that this Gaussian form fails to
reproduce the standard perturbative result for the $p_\perp$ spectrum
of the produced particles. In other words, while this logarithm is not
crucial in order to obtain a reasonable fit of small-$x$ DIS data, it
is crucial in order to get the correct spectrum shape in $pA$
collisions.}.  The $x$ dependence is reminiscent of the BFKL evolution
of the gluon structure function at small $x$. Then, in order to
determine the values of the three parameters $\sigma_0$, $\lambda$ and
$x_0$, they fitted the DIS data at HERA for $x<0.01$. The best fit was
obtained with $\sigma_0=23$mb, $\lambda=0.29$ and $x_0=3.10^{-4}$.

\subsubsection{Bartels-Golec-Biernat-Kowalski model - with evolution}
The previous model for the dipole cross-section was subsequently
improved in \cite{BarteGK1} in order to make the small dipole limit
agree with leading twist perturbative QCD, including the DGLAP
evolution of the gluon density. This new model improves the agreement
with the high $Q^2$ data (up to $Q^2\le 450$GeV${}^2$). In this
version the dipole-proton cross section is \cite{BarteGK1} (see also
\cite{Muell6}):
\begin{eqnarray}
\sigma_{\rm dipole}(\r_\perp,x) = \sigma_0 \Bigg[1- 
\exp\bigg\{-{\pi^2 r_\perp^2 \alpha_s (\mu^2) xG(x,\mu^2) \over 3 \sigma_0}\bigg\}
\Bigg]
\label{eq:bgbk}
\end{eqnarray}
where $\mu^2\equiv \mu_0^2 + C/r_\perp^2$ and all the parameters are
determined from fits to the DIS data from HERA. Using
Eq.~(\ref{eq:bgbk}) in (\ref{eq:DIScross-section}) gives the cross
section for the interaction of a virtual photon with a proton target
described by the Color Glass Condensate.  Since we are interested in
scattering off a nucleus target, we can replace $ xG(x,\mu^2)$ in
Eq.~(\ref{eq:bgbk}) by $A xG(x,\mu^2)$. The parameter $\sigma_0$
must also be scaled according to the transverse area of the target.

The goal of the rest of this paper is to show that once one knows the
dipole cross-section, there are several other interesting processes
for which one can make quantitative predictions, since they can be
expressed in terms of the same dipole cross-section. For instance, the
cross section for jet production in $pA$ collisions is very closely
related to the dipole cross-section on a nucleus, as we will see in
section \ref{sec:pA}.

\section{Virtual photon production in pA collisions}
\subsection{Photon production amplitude}
The first example we consider is the production of virtual photons
(or, equivalently, lepton pairs) in pA collisions, since this is in
fact the process which is the most closely related to DIS. We follow
here the description of the proton assumed in \cite{GelisJ1,GelisJ2},
where it was assumed that the proton can be described in terms of
standard parton distributions. Therefore, the relevant subprocess for
the production of a virtual photon in pA is $qA\to q\gamma^* X$, which
is related to $\gamma^* A\to q\bar{q}X$ by a crossing symmetry.

\begin{figure}[htbp]
\begin{center}
\resizebox*{!}{1.7cm}{\includegraphics{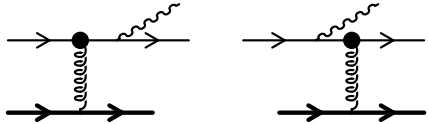}}
\end{center}
\caption{\label{fig:gamma-prod} The relevant diagrams 
  for $qA\to q\gamma^* X$ in the Color Glass Condensate model.  The
  black dots denote the eikonal interaction between the quark or
  antiquark with the classical color field.}
\end{figure}

The main difference in the case of $\gamma^*$ production is the fact
that there are only two diagrams, represented in figure \ref{fig:gamma-prod}.
Note that there could a priori be a third diagram with eikonal
interactions of the quark both before and after the photon emission.
However, this diagrams is in fact strongly suppressed by large energy
denominators \cite{GelisJ1}, which simply indicates that by the time
the photon has been emitted after a first scattering of the quark off
the nucleus, the nucleus is far behind and any further scattering is
impossible (in order words, the nucleus is so fast that the photon is
emitted outside of the nucleus).  This discrepancy between the number
of diagrams contributing to both processes is what makes the
correspondence between the two non-trivial\footnote{If we were working
at fixed order ${\cal O}(\alpha_s)$ instead of resumming all the
eikonal interactions, only the first two diagrams of figure
\ref{fig:DIS} would contribute to DIS, and finding the relation
between DIS and $\gamma^*$ production would be trivial. Note that a
similar correspondence was found in \cite{GelisP2}, although in a case
where the photon kinematics was far less general.}, and worth checking
explicitly.  In the CGC model, the $qA\to q\gamma^* X$
amplitude is given by\footnote{Again, we do not include the factor
$2\pi\delta(p^--q^--k^-)$ in the definition of ${\cal
M}^\mu_{\gamma^*}$.}
\begin{eqnarray}
&&{\cal M}^\mu_{\gamma^*}(\p|\q,\k)=i\int d^2\x_\perp
e^{i(\q_\perp+\k_\perp-\p_\perp)\cdot\x_\perp}
\left(U(\x_\perp)-1\right)\nonumber\\
&&\qquad\quad\times\overline{u}(\q)\left[
\frac{\gamma^\mu(\slQ+\slK+m)\gamma^-}{(Q+K)^2-m^2+i\epsilon}
+\frac{\gamma^-(\slP-\slK+m)\gamma^\mu}{(P-K)^2-m^2+i\epsilon}
\right]u(\p)\; ,
\label{eq:photon1}
\end{eqnarray}
where the two terms correspond to the two diagrams of figure
\ref{fig:gamma-prod} respectively. In this equation $p, q$ and $k$
are the momenta of the incoming quark, and of the outgoing quark and
photon respectively, and $\x_\perp$ is the transverse coordinate of the
quark. Using now the first of Eqs.~(\ref{eq:dirac-alg1}) as well as
\begin{equation}
u(\p)=\frac{1}{2p^-}(m+\slP)\gamma^- u(\p)\; ,
\end{equation}
and introducing again a dummy variable $\l_\perp$, we can after some
algebra rewrite the $\gamma^*$ production amplitude as:
\begin{eqnarray}
&&{\cal M}^\mu_{\gamma^*}(\p|\q,\k)=\frac{i}{2}
\int\frac{d^2\l_\perp}{(2\pi)^2}
\int d^2\x_{1\perp}d^2\x_{2\perp}
\nonumber\\
&&\qquad\;
\times e^{i\l_\perp\cdot\x_{1\perp}}
e^{i(\q_\perp+\k_\perp-p_\perp-l_\perp)\cdot\x_{2\perp}}
\left(U(\x_{1\perp})-U(\x_{2\perp})\right)\nonumber\\
&&\qquad
\;\times\overline{u}(\q)\,\Gamma^\mu(-k^\pm,-\k_\perp|q^-,-p^-,\q_\perp-\l_\perp)\,u(\p)\; ,
\label{eq:photon2}
\end{eqnarray}
where $\Gamma^\mu$ is again the function defined by
Eq.~(\ref{eq:commonspinors}).  We see that the only difference between
the DIS and $\gamma^*$-production amplitudes, besides the obvious
changes $P\to -P$ and $K\to -K$, is an extra factor
$U^\dagger(\x_{2\perp})$ under the integral. In other words, the two
amplitudes can be related by crossing symmetry up to a unitary matrix,
which violates a strict crossing symmetry at the amplitude level. This
has been noted independently by S.~Peign\'e in \cite{Peign2}, where
Drell-Yan and DIS are compared up to the terms involving three
scatterings on the target.  In the present case, this violation is in
fact due to the eikonal approximation. Indeed, this approximation
being valid for an asymptotically large relative momentum between the
projectile and the target, it does not allow to reverse continuously
the momentum of the projectile.

\subsection{Cross-section}
The photon production cross-section is obtained from the amplitude by
\begin{eqnarray}
&&
d\sigma_{\gamma^*}=\frac{d^3\q}{(2\pi)^22q_0}\frac{d^3\k}{(2\pi)^32k_0}
\frac{1}{2p^-} 
2\pi\delta(p^--q^--k^-)\nonumber\\
&&\qquad\qquad\times
\frac{1}{2N_c}
\left<{\cal M}^\mu_{\gamma^*}(\p|\q,\k)
{\cal M}^{\nu *}_{\gamma^*}(\p|\q,\k)
\right>_\rho
\epsilon_\mu(K)\epsilon_\nu^*(K)\; ,
\end{eqnarray}
where the factor $1/2N_c$ is for the average over the color and spin
of the incoming quark.  At this point, it is useful to replace the
integration variable $\l_\perp$ by $\m_\perp=\q_\perp-\l_\perp$ both
in the amplitude and in its complex conjugate, in order to exploit
the momentum dependence of the function $\Gamma^\mu$. Another
simplification comes from the fact that any term in the amplitude
squared depends non-trivially only on two out of the four transverse
coordinates. Therefore, the two transverse coordinates that appear
only in exponentials can be integrated out immediately, which gives
two delta constraints on some combination of transverse momenta, that
can be used in order to perform the integrals over $\m_\perp$ and
$\m^\prime_\perp$. Following this procedure, one obtains the following
form for the differential $\gamma^*$ production
cross-section\footnote{We also use the freedom to add a constant to
the term in $UU^\dagger$, because doing this does not change the final
result for the cross-section.}:
\begin{eqnarray}
&&d\sigma_{\gamma^*}=\frac{d^3\q}{(2\pi)^3 2q_0}
\frac{d^3\k}{(2\pi)^3 2k_0}
\frac{1}{2p^-} 
2\pi\delta(p^--q^--k^-)\,
\nonumber\\
&&\!\!\!\!\times
\int d^2\r_\perp e^{i(\q_\perp+\k_\perp-\p_\perp)\cdot\r_\perp}
\int d^2\X_\perp {\rm Tr}_c\left<
1-U(\X_\perp+\frac{\r_\perp}{2})
U^\dagger(\X_\perp-\frac{\r_\perp}{2})
\right>_\rho\nonumber\\
&&\!\!\!\!\times\frac{1}{2N_c}\Big[
M(\p_\perp-\k_\perp,\p_\perp-\k_\perp)
+M(\q_\perp,\q_\perp)\nonumber\\
&&\quad
-M(\q_\perp,\p_\perp-\k_\perp)
-M(\p_\perp-\k_\perp,\q_\perp)
\Big]\; ,
\label{eq:gamma-cs1}
\end{eqnarray}
where we use the shorthand
\begin{eqnarray}
&&M(\m_\perp,\m^\prime_\perp)\equiv
\epsilon_\mu(K)\epsilon^*_\nu(K)\;{\rm Tr}_d\big((\slq+m)
\Gamma^\mu(-k^\pm,-\k_\perp|q-,-p^-,\m_\perp)\nonumber\\
&&\qquad\qquad\qquad\qquad\times(\slp+m)
\Gamma^{\nu\dagger}(-k^\pm,-\k_\perp|q^-,-p^-,\m^\prime_\perp)
\big)\; .
\end{eqnarray}
In fact, the four terms in Eq.~(\ref{eq:gamma-cs1}) could have been
obtained directly by squaring the two terms of
Eq.~(\ref{eq:photon1})\footnote{The form of Eq.~(\ref{eq:photon2}) for
the $\gamma^*$ production amplitude is in fact not optimal in order to
calculate the cross-section because it contains some redundant
variables. Its only advantage is to make obvious the similarities with
the DIS amplitude.}. Again, we see that the dipole cross-section
appears naturally in this cross-section. Therefore, if one determines
the dipole cross-section from DIS data, then it becomes possible to
make quantitative {\sl predictions} for the production of virtual
photons in pA collisions. This possibility has already been exploited
in \cite{KopelRTJ1} in order to study nuclear effects on the Drell-Yan
process.

One can in fact obtain a form very similar to
Eq.~(\ref{eq:DIScross-section}) for the $\gamma^*$ production
cross-section integrated over the transverse momenta $\q_\perp$ and
$\k_\perp$ of the final particles:
\begin{equation}
\sigma_{\gamma^*}=\int_0^1 dz \int d^2\r_\perp
\left|\phi(p^-,\p_\perp|k^+,z,\r_\perp)\right|^2
\sigma_{\rm dipole}(\r_\perp)\; ,
\label{eq:gamma-cs2}
\end{equation}
where $z$ is the longitudinal momentum fraction taken by the photon
($k^-=zp^-$, $q^-=(1-z)p^-$) and where we denote
\begin{eqnarray}
&&\left|\phi(p^-,\p_\perp|k^+,z,\r_\perp)\right|^2\equiv
\frac{1}{64\pi p_-^2 z(1-z)}
\int\frac{d^2\q_\perp}{(2\pi)^2}
\int\frac{d^2\k_\perp}{(2\pi)^2}
e^{i(\q_\perp+\k_\perp-\p_\perp)\cdot\r_\perp}
\nonumber\\
&&\qquad\times\Big[
M(\p_\perp-\k_\perp,\p_\perp-\k_\perp)
+M(\q_\perp,\q_\perp)\nonumber\\
&&\qquad\quad
-M(\q_\perp,\p_\perp-\k_\perp)
-M(\p_\perp-\k_\perp,\q_\perp)
\Big]\; .
\end{eqnarray}

\section{Forward pion production in pA collisions}
\label{sec:pA}
\subsection{Quark scattering amplitude}
In this section we show how one can relate the cross section for
forward particle production in pA collisions to the dipole
cross-section. To be specific, we will consider pions but this can be
repeated for any particle for which there is a known fragmentation
function.  We will use the results of \cite{DumitJ1,DumitJ2} for
scattering of an on-shell quark from the target described by the Color
Glass Condensate.
\begin{figure}[htbp]
\begin{center}
\resizebox*{!}{1.3cm}{\includegraphics{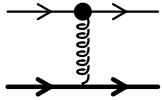}}
\end{center}
\caption{\label{fig:q-scatt} The relevant diagram 
  for $qA\to qX$ in the Color Glass Condensate model.  The
  black dots denote the eikonal interaction between the quark or
  antiquark with the classical color field.}
\end{figure}

The amplitude for scattering of a quark from the target described by a
classical field is given by the diagram represented in figure
\ref{fig:q-scatt}, and has been obtained in \cite{DumitJ1,DumitJ2}.
Factoring out the obvious $2\pi\delta(p^--q^-)$, this amplitude reads
(the same as Eq.~(\ref{eq:Teik})):
\begin{eqnarray}
{\cal M}_{qA\to qX}(\p|\q)=
\overline{u}(\q)
\left[\gamma^-\int d^2\x_\perp e^{i(\q_\perp-\p_\perp)\cdot \x_\perp}
(U(\x_\perp)-1)\right]u(\p)\; ,
\label{eq:qAampl}
\end{eqnarray}
where $\p$ and $\q$ are the momenta of the incoming and outgoing quark
respectively.

\subsection{Scattering cross section}
{}From this amplitude, it is easy to obtain the cross section
\cite{DumitJ1,DumitJ2}:
\begin{equation}
d\sigma_{qA\to qX}=\frac{d^3\q}{(2\pi)^3 2q_0}\frac{1}{2p^-} 
2\pi\delta(p^--q^-)
\frac{1}{2N_c}
\left<
{\cal M}_{qA\to qX}(\p|\q){\cal M}^*_{qA\to qX}(\p|\q)
\right>_\rho\; ,
\end{equation}
where the prefactor $1/2N_c$ comes from the average over the spin and
color of the incoming quark. This can be rewritten as:
\begin{eqnarray}
&&d\sigma_{qA\to qX}=\frac{d^2\q_\perp}{(2\pi)^2}
\frac{1}{N_c}\int d^2\r_\perp e^{i(\q_\perp-\p_\perp)\cdot\r_\perp}
\nonumber\\ &&\qquad\times \int d^2\X_\perp {\rm Tr}_c\left<
\left(U(\X_\perp+\frac{\r_\perp}{2})-1\right)
\left(U^\dagger(\X_\perp-\frac{\r_\perp}{2})-1\right) \right>_\rho\; .
\end{eqnarray}
 Note that for the scattering of a gluon, one would have to replace
the matrix $U$ by its analogue in the adjoint representation, but the
average over the color of the initial state would require a factor
$1/(N_c^2-1)$ instead of $1/N_c$.  Expanding the correlator in the
bracket $\left<\cdots\right>_\rho$, we can rewrite this cross-section
in terms of the dipole cross-section:
\begin{eqnarray}
&&d\sigma_{qA\to qX}=\frac{d^2\q_\perp}{(2\pi)^2}
\,dq^-\delta(q^--p^-)
\int d^2\r_\perp e^{i\q_\perp\cdot\r_\perp}
\nonumber\\ &&\!\!\!\!\times
\left[\frac{1}{N_c}
 \int d^2\X_\perp {\rm Tr}_c\left<
2-U(\X_\perp+\frac{\r_\perp}{2})
-U^\dagger(\X_\perp-\frac{\r_\perp}{2})
\right>_\rho
-\sigma_{\rm dipole}(\r_\perp)\right]
\; ,\nonumber\\
&&
\end{eqnarray}
where, anticipating the use of collinear factorization for the
incoming quark, we have set $\p_\perp=0$.  The first term
$\left<2-U-U^\dagger\right>_\rho$ is essential in order to obtain the
correct total cross-section. Indeed, integrating over $\q_\perp$
generates a $\delta(\r_\perp)$, and since $\sigma_{\rm
dipole}(\r_\perp=0)=0$, we have for the total $qA\to qX$
cross-section\footnote{In order for the optical theorem to hold, the
forward elastic ($qA\to qA$) amplitude must be:
\begin{eqnarray}
{\cal M}_{qA\to qA}^{\rm forward}=
2p^- \int d^2\x_\perp\left<U(\x_\perp)-1\right>_\rho\; .
\end{eqnarray}
In other words, the forward $qA\to qA$ amplitude is obtained from
Eq.~(\ref{eq:qAampl}) by setting $\p=\q$ and by averaging the
{\sl amplitude} over the classical color sources in the nucleus.}:
\begin{eqnarray}
\sigma_{qA\to qX}^{\rm total}=\frac{1}{N_c} \int d^2\X_\perp {\rm
 Tr}_c\left<2-U(\X_\perp) -U^\dagger(\X_\perp) \right>_\rho\; .
\end{eqnarray}

If we recall (see \cite{GelisP1} for instance) that
$\left<U(\x_\perp)\right>_\rho$ is suppressed exponentially like
$\exp(-Q_s^2/\Lambda_{_{QCD}}^2)$ when $Q_s\gg \Lambda_{_{QCD}}$, we
can approximate the $qA\to qX$ cross-section by:
\begin{eqnarray}
&&d\sigma_{qA\to qX}=\frac{d^2\q_\perp}{(2\pi)^2}\,dq^-\delta(q^--p^-)
\int d^2\r_\perp e^{i\q_\perp\cdot\r_\perp}
(2\pi R^2-\sigma_{\rm dipole}(\r_\perp))
\; ,\nonumber\\
&&
\label{eq:qAcs}
\end{eqnarray}
for a target of radius $R$. The term in $2 \pi R^2$ contributes only
to the scattering in the forward direction (i.e. at $\q_\perp=0$).

\subsection{Discussion}
\subsubsection{Cronin effect at the partonic level}
{}From the previous relations, it is trivial to write the non-forward
part of the $q_\perp$-spectrum as:
\begin{equation}
\frac{d\sigma_{qA\to qX}}{d^2\q_\perp}=-\frac{1}{(2\pi)^2}\,\widetilde{\sigma}_{\rm dipole}(q_\perp)\; ,
\label{eq:pA-cs}
\end{equation}
where $\widetilde{\sigma}_{\rm dipole}(q_\perp)$ is the Fourier
transform of the dipole cross-section, defined as:
\begin{equation}
\widetilde{\sigma}_{\rm dipole}(q_\perp)\equiv
\int d^2\r_\perp e^{i \q_\perp\cdot\r_\perp}\sigma_{\rm dipole}(\r_\perp)\; .
\end{equation}
In other words, the $q_\perp$-spectrum of particle production in $pA$
collisions is given by the opposite of the Fourier transform of the
dipole cross-section.

In the McLerran-Venugopalan model, this Fourier transform has been
studied in detail in \cite{GelisP1,GelisP2} where it controls the
production of $q\bar{q}$ pairs in ultra-peripheral $AA$ collisions and
was denoted $\pi R^2 C(\q_\perp)$ (see for instance the appendix B of
\cite{GelisP1}). In particular, the large $q_\perp$ behavior of this
function was found to be
\begin{equation}
C(\q_\perp)=
2{{Q_s^2}\over{q_\perp^4}}+{8\over\pi}{{Q_s^4}\over{q_\perp^6}}
\left(\ln\left({{q_\perp}\over{\Lambda_{_{QCD}}}}\right)-1\right)
+{\cal O}\left(
{{Q_s^6}\over{q_\perp^8}}\right)\; .
\label{eq:Ck-asympt}
\end{equation}
Note that if one replaces $Q_s^2$ by its expression in terms of parton
distributions inside the target, the first term in $1/q_\perp^4$
reproduces the perturbative QCD result for $qg\to qg$ scattering with
one gluon exchange in the t-channel (i.e. the dominant piece at high energy).
In addition, if one assumes that $Q_s^2$ scales like $A^{1/3}$ for
large nuclei, then the leading term of the $q_\perp$ spectrum simply
scales like $A$ (another $A^{2/3}$ comes from the transverse area $\pi
R^2$). If the second term in this expansion scales differently with
$A$, then there is a Cronin effect. From Eq.~(\ref{eq:Ck-asympt}), one
can see that the next-to-leading term in the $q_\perp$-spectrum is
scaling like $A^{4/3}$ and is positive. Therefore, one has a positive
Cronin effect in the McLerran-Venugopalan model. 
\begin{figure}
\centerline{\resizebox*{!}{7cm}{\rotatebox{-90}{\includegraphics{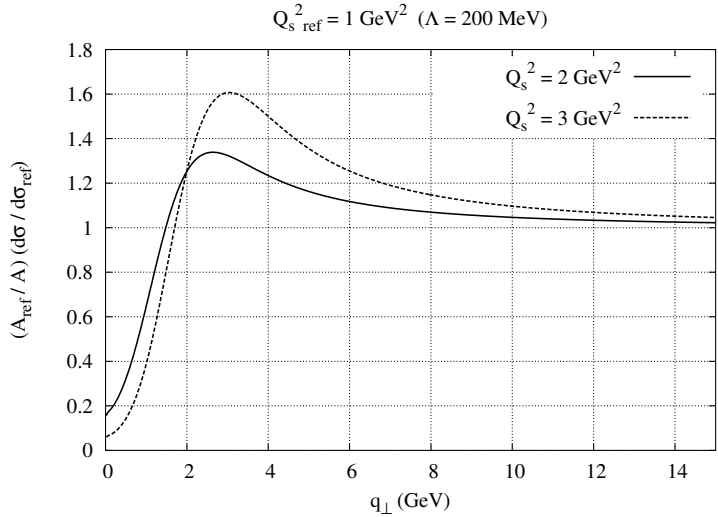}}}}
\caption{\label{fig:croninMV} Cronin effect at the partonic level 
  in the McLerran-Venugopalan model. The reference nucleus is chosen
  such that the corresponding saturation scale is $Q_s^2{}_{\rm
    ref}=1$GeV${}^2$. This reference value of $Q_s$ is compared to
  nuclei for which $Q_s^2=2$~GeV${}^2$ and $Q_s^2=3$~GeV${}^2$. The infrared cutoff is set to $\Lambda=200$~MeV.}
\end{figure}
The ratio $(A_{\rm
  ref}/A)d\sigma_{qA\to qX}/d\sigma_{qA_{\rm ref}\to qX}$, as
predicted in the McLerran-Venugopalan model, is plotted as a function
of $q_\perp$ in figure \ref{fig:croninMV}. The reference nucleus is
chosen such that the corresponding saturation scale is $Q_s^2{}_{\rm
  ref}=1~$GeV${}^2$, and the ratio is displayed for two nuclei such that the
saturation scales are $Q_s^2=2$~GeV${}^2$ and $Q_s^2=3$~GeV${}^2$.  One can see that this ratio goes to $1$ at large
transverse momentum, and exhibits a pronounced maximum at intermediate
$q_\perp$'s before dropping below $1$ at small transverse momenta.

Alternatively, one could take $A_{ref}=A=200$ and consider quark
nucleus scattering at mid rapidity, assuming that $Q_s^2{}_{\rm
ref}=1~$GeV${}^2$ at mid rapidity. Then the solid and dashed lines
correspond to the relative enhancement of the cross section in the
forward rapidity region, $y \sim 2.5$ and $y \sim 3.5$ units away from
mid rapidity (towards the proton), respectively.  The low $q_\perp$
spectrum is more suppressed as one goes to more forward rapidities and
high $q_\perp$ enhancement is stronger and moves to higher $q_\perp$.

It would be interesting to study whether this phenomenon is present in
more sophisticated models of the dipole cross-section, like the model
of Bartels, Golec-Biernat and Kowalski. It is also necessary to fold
this ``partonic level'' calculation with the proton parton
distribution functions, and with the appropriate fragmentation
functions, as this could potentially suppress the effect. A numerical
study of these issues is underway.

\subsubsection{Further constraints of the dipole cross-section in pA}
In DIS, one does not probe thoroughly the dipole cross-section, but
only its overlap with the square of the photon wave function. This
implies that some values of $r_\perp$ matter more than others in the
integral. One could therefore see Eq.~(\ref{eq:pA-cs}) as another way
of constraining the dipole cross-section. In particular, one would
like to recover the results of perturbative QCD at large $q_\perp$,
that is a cross-section that falls like $q_\perp^{-4}$ (up to
logarithms coming from the running $\alpha_s$ and from the DGLAP
evolution of the gluon distribution inside the target).

For instance, this remark pretty much excludes the
Golec-Biernat-W\"usthoff model of the dipole cross-section. Indeed,
in this model (see Eq.~(\ref{eq:dipGBW})) the Fourier transform that
gives the $q_\perp$-spectrum can be performed analytically, giving:
\begin{equation}
\frac{d\sigma_{qA\to qX}}{d^2\q_\perp}=\frac{R_0^2(x)\sigma_0}{\pi}\,
e^{-R_0^2(x) q_\perp^2} .
\end{equation}
The major difference compared to the result in the
McLerran-Venugo\-pa\-lan model is the large $q_\perp$ behavior, which
exhibits a Gaussian tail, as opposed to a power law tail. This fact
alone is probably enough to justify not considering this model any
further in order to study the region of large transverse momenta in pA
collisions.

\begin{figure}
\centerline{\resizebox*{!}{7cm}{\rotatebox{-90}{\includegraphics{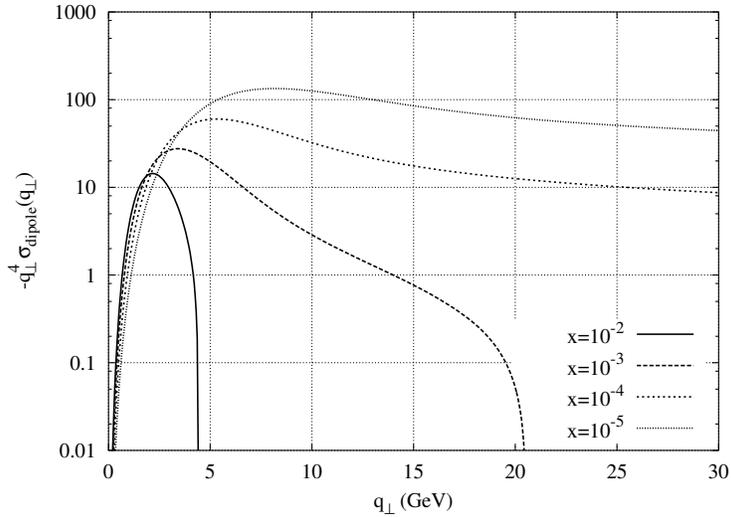}}}}
\caption{\label{fig:bartels-fourier} Values of $-q_\perp^4
\widetilde{\sigma}_{\rm dipole}(q_\perp)$ as a function of $q_\perp$,
for several values of $x$ in the model of Bartels, Golec-Biernat and
Kowalski.}
\end{figure}

For the Bartels-Golec-Biernat-Kowalski model, one needs to compute
numerically the Fourier transform of the dipole cross-section. In
order to do this, we have taken the parameters of their second fit
(``Fit 2'' in Table 1 of \cite{BarteGK1}), which lead to the results
displayed in figure \ref{fig:bartels-fourier}.
One can see that for very small values of $x$ ($x=10^{-4}, x=10^{-5}$
on this plot) one obtains the expected scaling at large
$q_\perp$. Indeed, $q_\perp^4$ times the Fourier transform of the dipole
cross section is almost constant. However, for larger values of $x$
($x=10^{-2}, x=10^{-3}$ on the plot), the Fourier transform of the
dipole cross-section suddenly drops at some $q_\perp$ and its sign
changes \footnote{This depends on whether a running or constant $\alpha_s$ 
is used \cite{HK}. For constant $\alpha_s$ one always gets a positive sign. 
See also eq. $20$ in \cite{BarteGK1} and the discussion afterwards.}. 

Above the $q_\perp$ where this happens, one would obtain a
negative $qA\to q X$ cross-section from formula Eq.~(\ref{eq:pA-cs}).
In this formula, one apparently probes different aspects of the dipole
cross-section compared to DIS, and the model by Bartels,
Golec-Biernat and Kowalski, although very successful at fitting all
the HERA DIS data at $x<10^{-2}$, does not give a consistent answer
for pA collisions at moderate $x$. Somehow, this defect is hidden in
DIS by the way the photon wave functions weights the different values
of $r_\perp$. Given this, one could perhaps reverse the main argument
of this paper, and consider pA collisions as a way of either fine
tuning the dipole model, or as a way of ruling it out.

\subsection{From $qA\to qX$ to $pA\to \pi X$}
To relate (\ref{eq:qAcs}) to proton-nucleus collisions, we can use
collinear factorization on the proton side and convolute
(\ref{eq:qAcs}) with the quark distribution function inside a proton
and a quark-pion fragmentation function to get the cross section for
$pA \rightarrow \pi X$.
\begin{eqnarray}
{d\sigma_{pA\to \pi(k)X} \over dk^-d^2\k_\perp}\equiv
\int dx_q\, dz\, q_p(x_q)\,\,
{d\sigma_{qA\to qX} \over dq^-d^2\q_\perp} \,\,
D_{q/\pi}(z)
\label{eq:fac}
\end{eqnarray}
where $q_p(x_q)$ is the distribution function of quarks with fractional
momentum $x_q$ inside a proton and $D_{q/\pi}(z)$ is the fragmentation
function of a quark into a pion carrying fraction $z$ of its energy. 
Using (\ref{eq:qAcs}) in (\ref{eq:fac}) we get
\begin{eqnarray}
&&{d\sigma_{pA\to \pi(y,\k_\perp)X} \over dyd^2\k_\perp}=
{1\over (2\pi)^2}\sqrt{k_\perp^2 \over s}\,e^y\,\int_{z_{\rm min}}^1 dz\,
q_p\left(x_q\right)\,D_{q/\pi}(z) \nonumber\\
&&\qquad\times 
\int d^2\r_\perp e^{i\k_\perp\cdot\r_\perp/z}
(2\pi R^2-\sigma_{\rm dipole}(\r_\perp,x_g))\; ,
\label{eq:csrap}
\end{eqnarray}
where we have used the following kinematical relations $x_q=k_\perp
e^y/z\sqrt{s}$, $x_g=k_\perp e^{-y}/z\sqrt{s}$, $z_{\rm min}= k_\perp
e^y/\sqrt{s}$ ($\sqrt{s}$ is the center of mass energy for a {\sl
proton-nucleon subsystem}). This way one can relate production of
pions, kaons, protons, neutrons, photons and dileptons in
proton-nucleus collisions in the forward rapidity region to the
structure functions measured in DIS. For this approach to be
self-consistent, one needs to make sure that the dominant contribution
to (\ref{eq:csrap}) comes from the small $x$ ($\le 0.01$) region. Our
preliminary result seem to indicate that this is indeed the case. A
detailed numerical study of particle production in proton-nucleus
collisions at RHIC is currently underway and will be reported
elsewhere.

\section{Conclusions}
In this paper, we have shown that in a classical description of the
gluon content of the nucleus, there are many interesting physical
quantities that can be related to the dipole cross-section (or, at a
more formal level, to the correlator
$\left<U(0)U^\dagger(\x_\perp)\right>$ of two Wilson lines). This was
already known for deep inelastic scattering and for the Drell-Yan
process when one considers cross-sections integrated over the
phase-space of the final state. In this paper, we establish also a
similar correspondence for forward particle production in proton
nucleus collisions, this time for the differential cross-section. More
precisely, we show that the $q_\perp$ spectrum of the produced
particles is related to the Fourier transform of the dipole
cross-section.

At a more formal level, we have generalized the standard leading twist 
collinear factorization approach to calculation of hadronic cross sections by 
showing that there is a universal object, the quark antiquark dipole 
cross section, which appears in all particle production cross sections
in high energy proton-nucleus collisions. This dipole cross section plays
the role of leading twist parton distributions in an all twist environment.
In principle, it can be measured in DIS or for example, in single inclusive 
pion production in high energy proton-nucleus collisions. Once it is measured
in a given process, it can be used to predict cross sections for other
processes such as single inclusive hadron, jet, photon or dilepton 
production in high energy proton-nucleus collisions.

This result could therefore be used in order to make predictions for
pA collisions, or to further constrain (or to rule out) the dipole
model with proton-nucleus experiments. The detailed numerical study
needed in order to make quantitative statements is in progress and
will be presented in a future work \cite{GelisJ3}. However, a preliminary
investigation of the Fourier transform of the dipole cross-section in
the model of \cite{BarteGK1} seem to lead to a negative $qA\to qX$
differential cross-section for some values of $x$ at large
$q_\perp$. This suggests that the small $r_\perp$ improvement of the
dipole cross-section made in \cite{BarteGK1} over
\cite{GolecW1,GolecW2} is still not quite able to ensure that the
results of perturbative QCD are recovered at large transverse momentum.

\vglue 5mm

\section*{\bf Acknowledgment}
We thank A. Dumitru, E.~Iancu, K.~Itakura, S.~Jeon, Y.~Kovchegov,
H.~Kowalski, S.~Peign\'e, D.~Teaney, R.~Venugopalan and W.~Vogelsang
for useful discussions.  We would also like to thank H.~Kowalski for
providing us with his code for calculating the dipole cross section in
the Bartels-Golec-Biernat-Kowalski model, and L.~Schoeffel for
providing us with his code of DGLAP evolution. J.J-M. is supported by
the U.S.\ Department of Energy under Contract No.\ DE-AC02-98CH10886
and in part by a PDF from BSA and would like to acknowledge the
hospitality of SUNY Stony Brook nuclear theory group while this work
was being completed. F.G. would like to thank the LPT/Orsay where part
of this work has been performed.

\bibliographystyle{unsrt}
%\bibliography{biblio}

\end{document}